\documentclass[fleqn,usenatbib]{mnras}

\usepackage{newtxtext,newtxmath}

\usepackage[T1]{fontenc}
\usepackage{ae,aecompl}

\usepackage[dvips]{color}
\usepackage{ulem}

\usepackage{graphicx}
\usepackage{lscape}
\usepackage{journals}

\newcommand{\flux}{erg s$^{-1}$ cm$^{-2}$}
\newcommand{\lum}{erg s$^{-1}$}
\newcommand{\nustar}{\textit{NuSTAR}}
\def\x8{M33~X-8}

\title[NuSTAR observations of ULX M33 X-8]{NuSTAR observations of the ultraluminous X-ray source M33 X-8: a black hole in a very high state?} 
\author[Krivonos et al.]{Roman\,Krivonos,$^{1}$\thanks{E-mail:
    krivonos@iki.rssi.ru (RK); sazonov@iki.rssi.ru (SS); sergey.tsygankov@utu.fi (SST); juri.poutanen@utu.fi (JP)} 
Sergey Sazonov,$^{1,2}$\footnotemark[1]  
Sergey S. Tsygankov,$^{1,3}$\footnotemark[1] 
and Juri Poutanen$^{1,3,4}$\footnotemark[1]\\
$^{1}$Space Research Institute of the Russian Academy of Sciences, Profsoyuznaya Str. 84/32, 117997 Moscow, Russia\\
$^{2}$National Research University Higher School of Economics, Myasnitskaya ul. 20, 101000 Moscow, Russia\\
$^{3}$Tuorla Observatory, Department of Physics and Astronomy,  FI-20014 University of Turku, Finland\\
$^{4}$Nordita, KTH Royal Institute of Technology and Stockholm University, Roslagstullsbacken 23, SE-10691 Stockholm, Sweden}

\begin{document}
\label{firstpage}
\pagerange{\pageref{firstpage}--\pageref{lastpage}}
\pubyear{2018}
\maketitle

\begin{abstract}
The closest known ultraluminous X-ray source (ULX), M33 X-8, has been recently observed with \nustar\ during its Extragalactic Legacy program, which includes a hard X-ray survey of the M33 galaxy. We present results of two long observations of M33 taken in 2017 March and July,  with M33 X-8 in the field of view. The source demonstrates a nearly constant flux during the observations, and its 3--20~keV spectrum can be well described by two distinct components: a standard accretion disc with a temperature of $\sim$1~keV at the inner radius and a power law with a photon index $\Gamma\approx 3$, which is significantly detected up to 20~keV. There is also an indication of a high-energy cutoff in the spectrum, corresponding to a temperature of the Comptonizing medium of $\gtrsim 10$~keV. The broad-band spectral properties of \x8\ resemble black hole X-ray binaries in their very high states, suggesting  that \x8\ is a black hole accreting at a nearly Eddington rate, in contrast to super-Eddington accretion believed to take place in more luminous ULXs. 
\end{abstract}


\section{Introduction}
\label{sec:intro}

The first detailed X-ray studies of the individual source content of ``normal'' Local Group galaxies (distance $\lesssim1$~Mpc) by the {\it Einstein} satellite have revealed a new class of intermediate luminosity ($L_{\rm X}\gtrsim 10^{39}$~\lum) X-ray sources \citep[cf.][]{1989ARA&A..27...87F}. Such objects, later found in larger numbers and at even higher ($\sim 10^{39}$--$10^{41}$~\lum) luminosities in more distant galaxies, are now widely referred to as ultraluminous X-ray sources \citep[ULX,][]{2000ApJ...535..632M}. Often, ULXs are defined as point-like off-nuclear sources whose apparent X-ray luminosity exceeds $10^{39}$~\lum, which roughly corresponds to the Eddington limit for a ``typical'' stellar-mass ($\sim10$~M$_{\sun}$) black hole \citep[e.g.,][]{1982AdSpR...2..177L,1989ARA&A..27...87F,1995ApJ...438..663M,1999ApJ...519...89C}.

The nature of ULXs has been under debate for a long time. Extensive observations carried out in the standard 2--10~keV X-ray band have provided increasing evidence that most ULXs are stellar-mass black holes accreting in super-Eddington regime \citep[see][for reviews]{2011NewAR..55..166F,2016AN....337..534R}. In this case, the accretion disc is expected to be slim in its inner region, expelling some mass in a wind and collimating the radiation along the symmetry axis \citep{1973A&A....24..337S,1999AstL...25..508L,2007MNRAS.377.1187P}. At large viewing angles, the central source is hidden by the optically thick wind and its high luminosity may be revealed only through its impact on the surroundings, via photoionization or dynamical action of the outflowing material. Such a scenario is likely applicable to the microquasar SS~433 in our Galaxy \citep{2004ASPRv..12....1F}. 

The discovery of periodic signals from a number of previously known ULXs \citep{2014Natur.514..202B,2016ApJ...831L..14F,2017MNRAS.466L..48I,2017Sci...355..817I} has provided strong evidence that a significant fraction of the ULX population may be powered by accretion onto a strongly magnetized neutron star. The disruption of the accretion disc by the stellar magnetic field at large radii can reduce the wind's collimation efficiency because the disc then remains in sub-critical accretion regime \citep{1982SvA....26...54L,2017MNRAS.470.2799C}. 
The presence of pulsations also argues against strong collimation, because reprocessing in the wind would smear the signal. This implies that the dominant source of radiation is then not the accretion disc but accretion columns \citep{2015MNRAS.454.2539M,2018MNRAS.476.2867M} or an envelope around the neutron star \citep{2017MNRAS.467.1202M}.

There is growing evidence that the spectra of black-hole ULX and ULX-pulsars are different. \cite{2018ApJ...856..128W} recently investigated the spectral signatures of the pulsed emission from three neutron-star ULXs, M82~X-2, NGC~7793~P13 and NGC~5907~ULX, finding that the accretion column (pulsed) emission in these sources dominates the total emission at high energies. Similar hard excesses are observed in a broader ULX sample with broadband coverage available to date, when the low-energy data are fitted with accretion disc models \citep[e.g.][]{2013ApJ...779..148W,2014ApJ...793...21W,2015ApJ...806...65W}. This suggests that a substantial fraction of the ULX population are neutron star accretors. We conclude that the broad-band spectroscopy becomes essential for understanding the nature of ULX central sources.

A salient ULX spectral feature, revealed by \textit{XMM--Newton}\ observations, is a turnover of the X-ray spectrum above $\sim$5--10~keV. This feature is not typical of normal (sub-Eddington) X-ray binaries \citep[e.g.][]{ZG04,2006csxs.book..157M,2007A&ARv..15....1D} and strongly suggests that super-Eddington accretion takes place in ULXs. However, as \textit{XMM--Newton}\ operates in the standard 2--10~keV X-ray band, observations of ULXs above 10 keV were highly anticipated to better constrain their spectra. The first attempt to perform imaging hard X-ray observations of ULXs, namely of M82 X-1 and Ho IX X-1, was undertaken by the {\it INTEGRAL} observatory \citep{2003A&A...411L...1W}. These observations revealed a rollover above $\sim$10~keV in the spectra of both sources \citep{2014AstL...40...65S}. 
This fact was later confirmed by the {\it Nuclear Spectroscopic Telescope Array} (\nustar) \citep{2016ApJ...829...28B,2014ApJ...793...21W,2017ApJ...839..105W}. 
\nustar, with its broad (3--79~keV) energy response, has opened a new era of hard X-ray observations of ULXs, providing a large set of broad-band ULX spectra \citep{2013ApJ...779..148W,2015ApJ...799..122W,2015ApJ...806...65W,2018MNRAS.473.4360W,2015ApJ...808...64M,2015ApJ...799..121R,2015ApJ...815...36A,2016MNRAS.463..756K,2017ApJ...839...46S}.

\x8\ \citep{1981ApJ...246L..61L,1988ApJ...325..531T} is the closest known ULX. It has an X-ray luminosity of a few $10^{39}$\lum\ and is located in the nearby spiral galaxy M33 (=NGC 598) at a distance of $\sim$817~kpc \citep[e.g.,][]{2011ApJS..193...31T}. Although the position of the X-ray source is consistent with the center of M33 \citep{2002MNRAS.336..901D,2004A&A...425...95D}, no activity in the nucleus of the galaxy has been revealed at other wavelengths. This testifies against an active galactic nucleus nature of \x8. Furthermore, photometry and kinematics measurements of the nuclear region of M33 indicate that its central black hole, if any, has a mass less that 1500~M$_{\sun}$ \citep{1993AJ....105.1793K,2001Sci...293.1116M,2001AJ....122.2469G}. Finally, the discovery of the 106 days periodicity of \x8\ \citep{1997ApJ...490L..47D} strongly contradicts the active galactic nucleus hypothesis and adds further evidence for an X-ray binary system, where modulations can be caused by precession of the accretion disc \citep[e.g.,][]{1996ApJ...472..582M}. Note also that similar long-term periodicities have been recently detected in some ULX-pulsars  (e.g.\citealt{2016ApJ...827L..13W}).

\x8\ has been the subject of many studies \citep{1987A&A...175...45G,1988ApJ...325..531T,1994ApJ...436L..47T,2004A&A...416..529F,2009PASJ...61.1287W,2011MNRAS.417..464M,2012PASJ...64..119I,2015A&A...580A..71L,2017ApJ...836...48S}. Most of the observations of the source were carried out at energies below 10~keV. The spectra measured in the 2--10~keV band appear to be dominated by thermal emission from an optically thick, geometrically thin or slim accretion disc with a temperature of $\sim$1~keV and allow for the presence of an additional power-law component with a photon index $\Gamma\approx 2$. The 2--10~keV spectrum of \x8\ thus resembles both the spectra of some other ULXs and those of Galactic black-hole binaries in their high and very high states. However, the narrowness of the energy band used for studying \x8\ so far allows for a lot of freedom in spectral modelling. It is thanks to the broad-band (3--79~keV) energy coverage of  \nustar\ that we are able to reliably constrain the spectral components of \x8\ for the first time.

\section{Observations and data analysis}
\label{sec:data}

\subsection{\nustar}

The ULX \x8\ was observed with \nustar\ \citep{nustar} in 2017 during a hard X-ray survey of the nearby spiral galaxy M33 performed as part of the Extragalactic Legacy program.\footnote{\url{https://www.nustar.caltech.edu/page/legacy_surveys}} 
\nustar\ carries two identical co-aligned X-ray telescopes with angular resolution of 18\arcsec\ (full width at half maximum, FWHM). The focal plane detector units of each telescope, referred to as focal plane modules A and B (FPMA and FPMB), cover a wide energy band 3--79~keV and provide spectral resolution of 400~eV (FWHM) at 10 keV.

\begin{figure}
\includegraphics[width=\columnwidth]{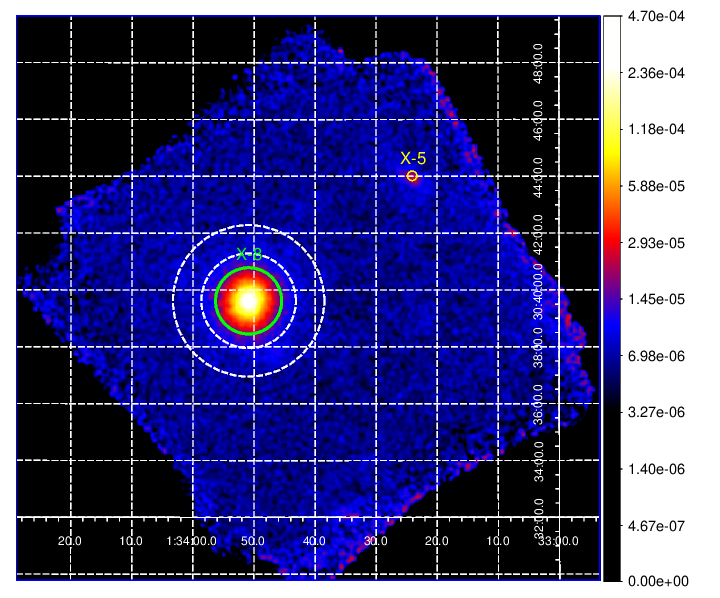}
\caption{\label{fig:nustar:m33} The \nustar\ count-rate image of the bulge of the M33 galaxy in the 3--79~keV energy band, with the total exposure of 207~ks. The solid green circle of radius $R=70\arcsec$ denotes the region used for source spectrum extraction. The dashed annulus around it, limited by radii $R=100\arcsec-160\arcsec$, shows the corresponding background region. The yellow region shows the position of the X-ray source M33 X-5 detected in the survey of M33 with \textit{ROSAT} \citep{1995ApJ...441..568S}.
}
\end{figure}

The \nustar\ M33 survey consists of observations of the central region and the disc of the galaxy, carried out in two campaigns in 2017 March and July. We selected the central parts of the survey, listed in Table~\ref{tab:log}, where \x8\ was located within the \nustar\ FOV. The total exposure of the data set is 207~ks. Fig.~\ref{fig:nustar:m33} shows a mosaic sky map obtained by stacking the FPMA and FPMB images in the full 3--79~keV bandpass.

\begin{table}
\noindent
\centering
\caption{List of \nustar\ observations of \x8.}\label{tab:log} 
\centering
\vspace{1mm}
  \begin{tabular}{|c|c|r|c|c|c|c|}
\hline\hline
 Date & ObsID & Exp. & Net count rate & Total count\\
 & & (ks) & count~s$^{-1}$ & \\
\hline
 2017-03-06 & 50310002001 & 105.5 & 0.16 & 33800\\
 2017-07-23 & 50310002003 & 101.8  & 0.14 & 28500\\
\hline
\end{tabular}\\
\begin{flushleft}
  {\it Note.} The total number of counts is given as a sum of those detected by FPMA and FPMB.
\end{flushleft}
\vspace{3mm}
\end{table}
 
\begin{figure}
\includegraphics[width=\columnwidth]{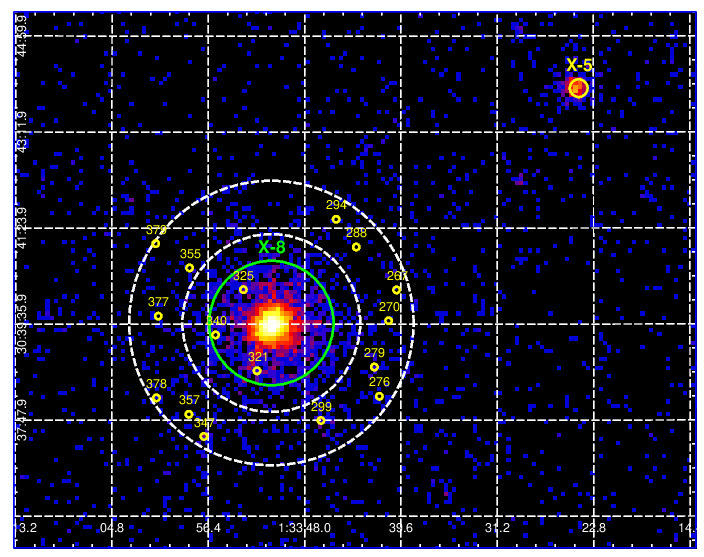}
\caption{Same as Fig.~\ref{fig:nustar:m33}, but for the \textit{Swift}-XRT full-band (0.2--10~keV) photon counting image of \x8. Yellow labels show the positions of X-ray sources detected in the \textit{Chandra} survey of M33 \citep{2011ApJS..193...31T} within the \nustar\ source and background spectral extraction regions. 
}\label{fig:xrt}
\end{figure}

Because \x8\ produces a relatively high count rate of $\sim$0.15~count~s$^{-1}$ on the \nustar\ detectors, we extracted the spectrum of the source from a large circular region with radius $R=70\arcsec$, enclosing $\sim$80 per cent of the point spread function \citep{2015ApJS..220....8M}, centered at the position of M33 X-8 \citep[RA=01:33:50.89, Dec=+30:39:37.2, J2000,][]{2004A&A...416..529F}. The \nustar\ spectrum was extracted using the {\sc nuproducts} task of the \nustar\ Data Analysis Software ({\sc nustardas}) v.1.8.0 and {\sc heasoft} v6.22.1.  The \nustar\ count rate within the $70\arcsec$ region is fully dominated by \x8. Nevertheless, it can contain contributions from other X-ray sources located within $\sim277$~pc from \x8\ (projected distance corresponding to $70\arcsec$). The \textit{Chandra} survey of M33 \citep{2011ApJS..193...31T} revealed three sources (SrcID. 321, 325 and 340) in the nuclear region within $70\arcsec$ from \x8 (see Fig.~\ref{fig:xrt}). Their total flux in the 0.35--8.0~keV band ($9.1\times10^{-7}$~phot~s$^{-1}$~cm$^{-2}$) is $\sim2\times10^{-4}$ times that of \x8\ ($4.3\times10^{-3}$~phot~s$^{-1}$~cm$^{-2}$, SrcID. 318). The background spectrum was extracted from the annulus region between $R=100\arcsec$ and $R=160\arcsec$ (Fig.~\ref{fig:nustar:m33}). Within this region, there are 13 \textit{Chandra} sources in the list of \citet{2011ApJS..193...31T}, with the total flux of $4.1\times10^{-5}$~phot~s$^{-1}$~cm$^{-2}$ in the 0.35--8.0~keV band, which corresponds to $\sim3\times10^{-3}$ of the \x8\ flux (properly scaled to the area of the source region). We conclude that weak sources both within the \nustar\ source and background extraction regions do not significantly contaminate the measured flux of \x8\ in the 0.35--8.0~keV band and assume the same to be the case at higher energies.

\subsection{{\it Swift}/XRT}

The \textit{Neil Gehrels Swift Observatory} ({\it Swift}) \citep{swift} has explored the central part of the M33 galaxy many times, targeting \x8\ \citep{2015A&A...580A..71L}. We utilized data from the XRT telescope \citep{2005SSRv..120..165B} obtained during the long ($\sim$20~ks) on-axis observation of \x8\ in Photon Counting (PC) mode on 2012 November 5 (ObsID. 00031042002).  It is the closest in time to the \nustar\ observations and of the best quality in terms of off-axis distance and exposure. The spectrum extraction was done using the online tools provided by the UK Swift Science Data Centre\footnote{\url{http://www.swift.ac.uk}}. The pile-up issues of \x8\ noticed by \citet{2015A&A...580A..71L} have been  resolved by the online spectrum extraction software \citep{2009MNRAS.397.1177E}. In particular, the code identifies those time intervals when the count rate is above 0.6 count s$^{-1}$ within a 30 pixel radius around the source, builds the PSF profile of the source and compares it with the calibrated, non-piled-up PSF \citep{2005SPIE.5898..360M}. If the source is piled up, the algorithm excludes the inner PSF region affected by pile-up from the spectral extraction.

\section{Spectral analysis}
\label{sec:spec}
\subsection{General procedure}

The spectral analysis was done using the X-ray Spectral Fitting Package \citep[{\sc xspec}][]{xspec}, version 12.9.1, part of the HEASOFT software package (version 6.22.1). The fitting procedures were conducted with spectra grouped to a minimum of 30 counts per bin to allow the use of $\chi^2$ statistics. We quote errors at 90 per cent confidence intervals unless stated otherwise.

Each spectral model discussed below has an absorption component \citep[{\sc phabs}), with the abundances adopted from][]{1989GeCoA..53..197A} and the absorbing line-of-sight column density fixed at the Galactic value $N_{\rm H}=1.1\times10^{21}$~cm$^{-2}$ \citep{2005A&A...440..775K}. An intrinsic absorption, if required by the fitting procedure, is included as a second  {\sc phabs} component with a free column density parameter.

\begin{table}
\noindent
\centering
\caption{Best-fit parameters for a power-law model obtained for both epochs of the \nustar\ observations in the 3--20~keV energy band.}\label{tab:pow}
\centering
 \begin{tabular}{|c|c|r|r|r|c|c|}
\hline\hline
Parameter & Units &  Epoch 1 &  Epoch 2 \\
\hline
\multicolumn{4}{c|}{{\sc const$\times$powerlaw}}\\
\hline
Constant$^{a}$& & $1.08\pm0.02$ & $1.01\pm0.02$ \\
$\Gamma$& & $3.29\pm0.03$  & $3.36\pm0.03$  \\
$N_{\rm pow}$& $10^{-2}$@1~keV & $2.69\pm0.12$  & $2.92\pm0.14$  \\
$F_{\rm 3-20\ keV}^{\rm pow}$& $10^{-12}$\flux & $7.36\pm0.11$  &$7.11\pm0.12$\\
$\chi^{2}$/dof &   & 483/386  &  453/352  \\
\hline
\end{tabular}\\
\begin{flushleft}
  $^{a}$The cross-calibration constant term is fixed at unity for FPMA and fitted for FPMB.
\end{flushleft}
\end{table}

\begin{figure}
\includegraphics[width=\columnwidth]{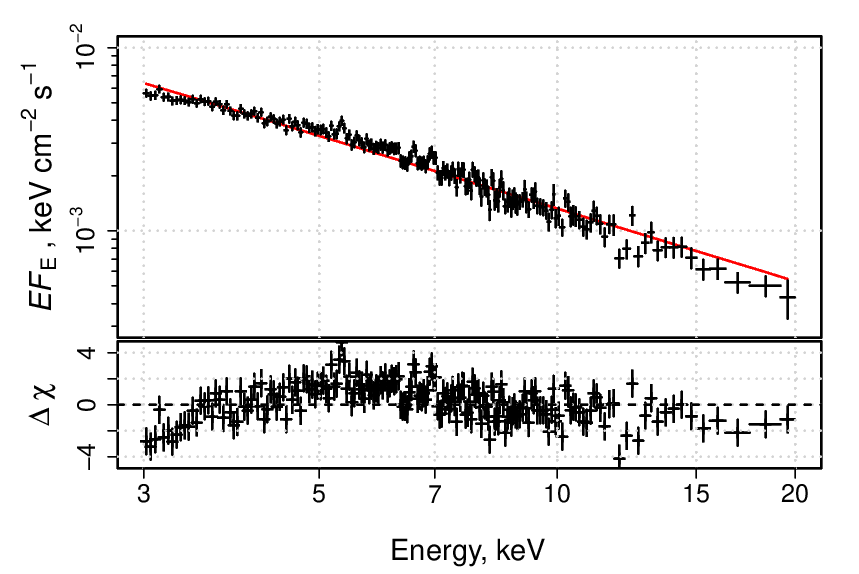}
\caption{Average \nustar\ X-ray spectrum of \x8\ obtained during the observations in 2017 (black crosses) and approximated by a power law (red line) with a  photon index $\Gamma=3.32\pm0.02$ and normalization $N_{\rm pow}=(2.8\pm0.1)\times10^{-2}$, giving $\chi^{2}$/dof = 924/739. }\label{fig:nustar:pow}
\end{figure}

To estimate possible variability of \x8\ between the two epochs of \nustar\ observations (see Table~\ref{tab:log}), we fitted the 3--20~keV spectra by a simple power-law model without any intrinsic absorption, as summarized in Table~\ref{tab:pow}. A cross-calibration coefficient between the FPMA and FPMB detectors was included in the fit and indicated that the FPMB spectrum had an 8 per cent higher normalization than the FPMA one in Epoch 1. This difference is somewhat larger than expected \citep[0--5 per cent,][]{2015ApJS..220....8M}. The epoch 2 FPMA/FPMB spectra have almost identical normalizations (with a difference of $\sim$1 per cent). We fixed the relative normalization between the \nustar\ modules at these values for the following analysis. We also truncated the \nustar\ spectra above 20~keV, where the source spectrum is dominated by the background. As seen from Table~\ref{tab:pow}, the best fitting parameters of the power-law model indicate low variability between the two epochs. The slope of the epoch 2 spectrum is marginally steeper, and there is a slight decrease in the 3--20~keV flux. We then combined both epochs into one data set to improve the statistics and to obtain a time-averaged broad-band spectrum of \x8. Note that we did not co-add spectra and response files according to the \nustar\ Help Desk recommendations\footnote{\url{https://heasarc.gsfc.nasa.gov/docs/nustar/nustar_faq.html}}, but rather fitted the individual spectra simultaneously. We stack \nustar\ spectra to one group for plotting purposes only.

The {\it Swift}/XRT long observation of \x8\ is used to extend the broad-band spectrum to energies below 5~keV, where the effective area of \nustar\ drops rapidly \citep{2015ApJS..220....8M}. In order to take into account possible source flux variations, we introduced a cross-normalization constant between the {\it Swift}/XRT and \nustar\ data.

\subsection{Simple models}

We first fitted the \nustar\ spectrum alone using a simple power-law model, as shown in Fig.~\ref{fig:nustar:pow}. The fit is poor with $\chi^2 = 924$ for 739 degrees of freedom (dof). The best fitting spectral index $\Gamma=3.32\pm0.02$ is consistent with that reported in Table~\ref{tab:pow} for both epochs of the \nustar\ observations. A significant curvature is evident over the 3--20 keV band, as shown in the lower panel of Fig.~\ref{fig:nustar:pow}.

\begin{figure}
\includegraphics[width=\columnwidth]{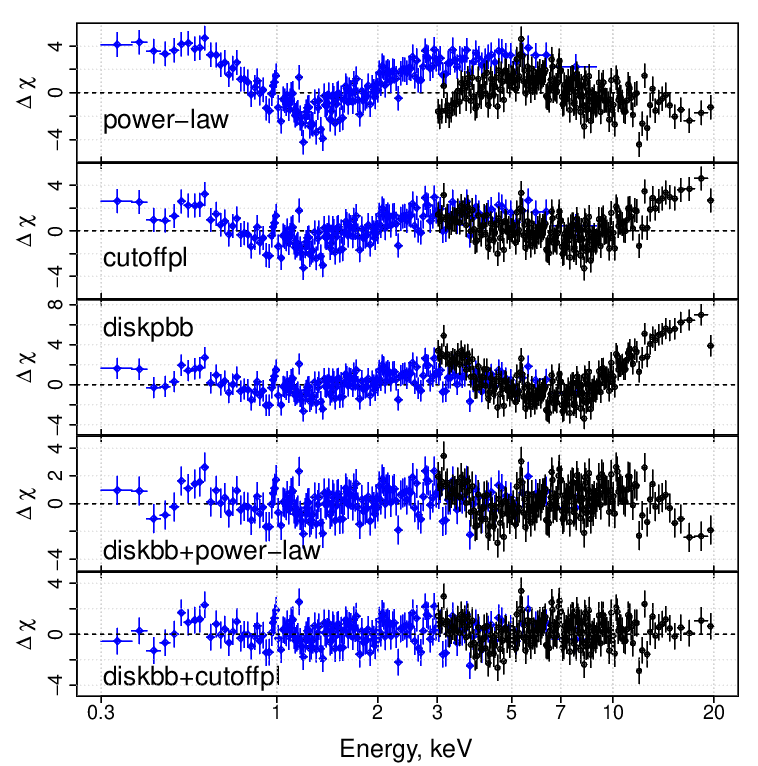}
\caption{Residuals for the phenomenological spectral models considered in this study (from top to bottom): a power law, a power law with an exponential high-energy cut-off, multi-colour disc with the $r^{-p}$ temperature profile, a combination of a standard accretion disc and a power law, and a combination of a standard accretion disc and a power law with an exponential high-energy cutoff. Blue diamonds and black circles show the {\it Swift}/XRT and \nustar\ data points, respectively.}
  \label{fig:simple1}
\end{figure}

We next fitted the broad-band 0.3--20~keV {\it Swift}/XRT and \nustar\ data using a number of phenomenological models. We start with simple continuum models modified by additional intrinsic absorption if required by the fit. The list of the models contains: (1) a power-law model; (2) a power law with an exponential cutoff ({\sc cutoffpl}); (3) a multi-colour disc model with the radial temperature dependence $r^{-p}$, with $p$ being a free parameter
\citep[{\sc diskpbb},][]{1994ApJ...426..308M}; (4) a power law combined with a standard multi-colour disc model \citep[{\sc diskbb},][]{1973A&A....24..337S}; (5) a power law with an exponential cutoff ({\sc  cutoffpl}) combined with {\sc diskbb}. Table~\ref{tab:simple} lists the best fitting parameters and Fig.~\ref{fig:simple1} shows the statistical residuals for each model.

\begin{table}
\noindent
\centering
\caption{Best fitting parameters for a number of phenomenological models of the broad-band 0.3--20~keV {\it Swift}/XRT and \nustar\ spectrum.}\label{tab:simple}
\centering
\vspace{1mm}
 \begin{tabular}{|c|c|r|r|r|c|c|}
\hline\hline
Parameter & Units &  Value \\
\hline
\multicolumn{3}{c|}{Model 1: {\sc phabs$\times$const$\times$power-law}}\\
\hline
$N_{\rm H}$ & $10^{22}$~cm$^{-2}$ &  $0.59\pm0.04$ \\
Constant$^a$ & & $0.52\pm0.03$ \\
$\Gamma$& & $3.32\pm0.02$  \\
$N_{\rm pow}$& $10^{-2}$@1~keV & $2.9\pm0.1$  \\
$\chi^{2}$/dof& & 1690/903 \\

\hline
\multicolumn{3}{c|}{Model 2: {\sc phabs$\times$const$\times$cutoffpl}}\\
\hline
$N_{\rm H}$ & $10^{22}$~cm$^{-2}$ &  $0.25\pm0.03$ \\
Constant & & $0.80\pm0.04$ \\
$\Gamma$& & $2.0\pm0.1$  \\
$E_{\rm fold}$& keV & $4.7\pm0.4$  \\
$N_{\rm cut}$& $10^{-2}$@1~keV & $1.0\pm0.1$  \\
$\chi^{2}$/dof& & 1207/902 \\

\hline
\multicolumn{3}{c|}{Model 3: {\sc phabs$\times$const$\times$diskpbb}}\\
\hline
$N_{\rm H}$ & $10^{22}$~cm$^{-2}$ &  $0.18\pm0.01$ \\
Constant & & $0.99\pm0.04$ \\
$kT$ & keV &   $1.88\pm0.02$ \\
$p$&   & 0.5 (hard limit)  \\
$N_{\rm disk}$ & $10^{-2}$ & $2.0\pm0.1$ \\
$R\cos^{1/2}i$ & km & $11\pm1$ \\
$\chi^{2}$/dof& & 1584/902 \\

\hline
\multicolumn{3}{c|}{Model 4: {\sc phabs$\times$const$\times$(diskbb+powerlaw)}}\\
\hline
$N_{\rm H}$ & $10^{22}$~cm$^{-2}$ &  $0.17\pm0.03$ \\
Constant & & $1.00\pm0.04$ \\
$kT$ & keV &   $1.21\pm0.02$ \\
$N_{\rm disk}$ & & $0.21\pm0.02$ \\
$R\cos^{1/2}i$ & km & $37\pm2$ \\
$\Gamma$& & $2.54\pm0.08$  \\
$N_{\rm pow}$& $10^{-3}$@1~keV & $3.1\pm0.6$  \\
$\chi^{2}$/dof& & 951/901 \\

\hline
\multicolumn{3}{c|}{Model 5: {\sc phabs$\times$const$\times$(diskbb+cutoffpl)}}\\
\hline
$N_{\rm H}$ & $10^{22}$~cm$^{-2}$ &  $<5^{a}$ \\
Constant & & $1.03\pm0.04$\\
$kT$ & keV &   $1.05\pm0.03$\\
$N_{\rm disk}$ & & $0.50\pm0.08$\\
$R\cos^{1/2}i$ & km & $58\pm5$\\
$\Gamma$& & $1.0\pm0.3$\\
$E_{\rm fold}$& keV & $5.3\pm0.4$  \\
$N_{\rm cut}$& $10^{-4}$@1~keV &  $7\pm3$ \\
$\chi^{2}$/dof& & 891/900 \\
\hline
\end{tabular}
\begin{flushleft}
  $^a$ The 2$\sigma$ upper limit.
\end{flushleft}
\end{table}

The power-law fit for the broad-band spectrum demonstrates strong curvature variations over the 0.3--20~keV energy band, as shown in the upper panel of Fig.~\ref{fig:simple1}. The quality of the fit is poor ($\chi^{2}$/dof = 1690/903), although the power-law parameters are generally consistent with those derived for the \nustar\ data set alone (Fig.~\ref{fig:nustar:pow}). The second model, with a power law modified by a high-energy exponential cutoff substantially improves the fit to $\chi^{2}$/dof = 1207/902, revealing an energy rollover at $\sim$5~keV. This model, however, still has significant deviations from the data over the considered energy band.

The third model, a modified version of the multi-colour disc model, is often used in the literature to describe ULX spectra, including \x8\ \citep[e.g.][]{2012PASJ...64..119I}. This single-component model does not fit the broad-band data ($\chi^{2}$/dof = 1584/902), leaving a strong excess at energies above 10~keV, as shown in the middle panel of Fig.~\ref{fig:simple1}. Similar high-energy tails seen in \nustar\ observations of other ULXs \citep{2013ApJ...779..148W,2014ApJ...793...21W,2015ApJ...806...65W,2015ApJ...808...64M} are usually attributed to Comptonization of accretion disc photons by a hot corona. 

The fourth model, a combination of a power law and an accretion disc component {\sc diskbb} significantly improves the fit ($\chi^{2}$/dof = 951/901) revealing a power-law-like excess with $\Gamma=2.54\pm0.08$ above 10~keV. However, the fitting residuals still demonstrate substantial variations, as evident in Fig.~\ref{fig:simple1}. The power-law photon index $\Gamma$ and the inner disc temperature are consistent with the results of previous studies of \x8, where a similar spectral model and data from {\it Swift}/XRT \citep{2015A&A...580A..71L}, {\it XMM-Newton} \citep{2004A&A...416..529F}, or {\it Suzaku} \citep{2012PASJ...64..119I} were used. The innermost disc radius, $R_{\rm in}$, can be directly estimated from the normalization of the {\sc diskbb} component: $N_{\rm disc}= R^2_{\rm in}\cos i/D^{2}$, where $D$ is the distance to the source in units of 10 kpc and $i$ is the inclination of the disc. The estimated inner disc radius, $R_{\rm in}=(37\pm2)\cos^{-1/2} i$~km, is in good agreement with the results of previous studies \citep[e.g.,][]{2011MNRAS.417..464M}. Associating this size with the radius of the innermost stable orbit of a standard accretion disc suggests that the relativistic compact object has a mass of at least $\sim 4$~$M_{\sun}$, with the exact value depending on the (unknown) inclination of the disc. More accurate estimation should take into account the inner boundary condition, the spectral hardening and relativistic corrections, which in turn would depend on the structure of the accretion disc (whether it is thin or slim). Nevertheless, the above crude estimate is consistent with the compact object being a stellar-mass black hole. 

Replacing the power-law component by its modification with an exponential cutoff (Model 5) further improves the fit statistics to $\chi^{2}$/dof =891/900, indicating the presence of a high-energy cutoff in the spectrum (the F-test probability is $10^{-14}$ for the null-hypothesis Model~4).

We conclude that the broad-band 0.3--20~keV spectrum of \x8\ cannot be described by a single accretion disc model {\sc diskpbb}, which was previously used to successfully fit some narrow-band spectra of this source \citep[e.g.][]{2015A&A...580A..71L}. The broad-band spectrum strongly requires an additional power-law component with a possible high-energy cutoff.

\subsection{Physically motivated models}

We have demonstrated that the broad-band spectrum of \x8\ certainly contains two components: the first one is well approximated by a multi-colour disc with the inner temperature of $kT\sim1$~keV, and the second one is a power-law-like high energy tail detected up to 20~keV. As often discussed in the literature \citep[e.g.][]{1999MNRAS.309..496G,2009PASP..121.1279S}, the combination of a low-energy blackbody-like component with a high-energy power law does not allow one to properly model the spectrum at low energies, because a steep power law extending to the soft part of the spectrum requires the introduction of an artificially high absorption column density and can result in erroneous determination of the disc properties.

\begin{table}
\noindent
\centering
\caption{Same as Table~\ref{tab:simple}, but for Models 3a and 3b.}\label{tab:model3}
\centering
\vspace{1mm}
 \begin{tabular}{|c|c|r|r|r|c|c|}
\hline\hline
Parameter & Units &  Value \\
\hline
\multicolumn{3}{c|}{Model 3a: {\sc const$\times$simpl$\otimes$diskpbb}}\\
\hline
Constant & & $0.87\pm0.04$ \\
$kT$ & keV &   $0.9\pm0.1$ \\
$p$& & $0.77\pm0.03$  \\
$N_{\rm diskpbb}$ &  & $1.4\pm0.4$ \\
$R\cos^{1/2}i$  & km & $97\pm13$ \\
$\Gamma$& & $3.4\pm0.1$  \\
$f_{\rm scat}$& & $0.43\pm0.06$  \\
$\chi^{2}$/dof & & 901/901 \\
\hline
\multicolumn{3}{c|}{Model 3b: {\sc const$\times$(simpl$\otimes$diskpbb)$\times$spexpcut}}  \\
\hline
Constant & & $0.72\pm0.06$ \\
$kT$ & keV &   $1.12\pm0.03$ \\
$p$& & $0.76_{-0.01}^{+0.03}$  \\
$N_{\rm diskpbb}$ &  & $0.93_{-0.12}^{+0.20}$ \\
$R\cos^{1/2}i$  & km & $79\pm10$ \\
$\Gamma$ & & $1.8\pm0.1$  \\
$f_{\rm scat}$& & $0.37\pm0.03$  \\
$E_{\rm cut}$& keV& $7.5_{-0.5}^{+0.8}$  \\
$\chi^{2}$/dof& & 891/900 \\
\hline
\end{tabular}
\end{table}

\begin{figure}
\includegraphics[width=\columnwidth]{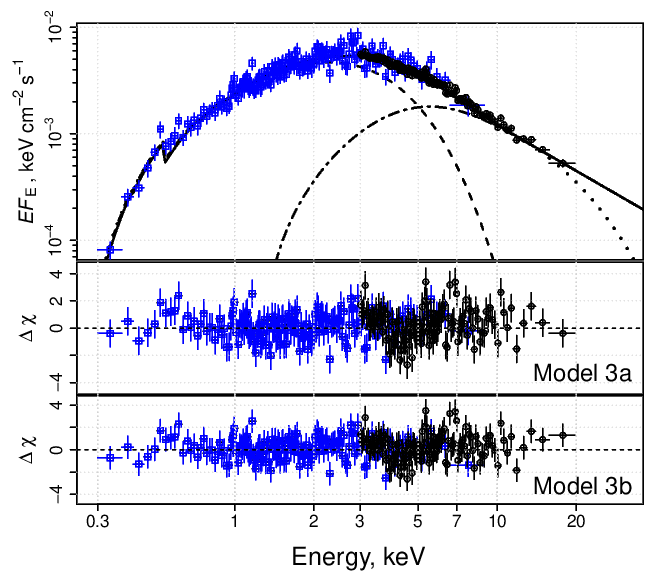}
\caption{Broad-band {\it Swift}/XRT and \nustar\ spectrum of \x8, shown by blue squares and black circles, respectively. The best-fitting spectral Model~3a ({\sc simpl$\otimes$diskpbb}, Table~\ref{tab:model3}) is shown by the solid black curve, along with a single accretion disc component {\sc diskpbb} ($kT=0.9\pm0.1$~keV, dashed curve) and the difference between the latter and the total model spectrum (i.e. {\sc simpl$\otimes$diskpbb}$-${\sc diskpbb}, the dot-dashed curve). The dotted curve represents the best-fitting Model~3b {\sc (simpl$\otimes$diskpbb)$\times$spexpcut}. The two bottom panels show the fit residuals for the respective models.}\label{fig:model3ab}
\end{figure}

This problem can be solved by using a self-consistent Comptonization model that has a low-energy cutoff in the otherwise power-law-like spectrum, such as the convolution model {\sc simpl} \citep{2009PASP..121.1279S}, which provides a description of Comptonization based on the non-relativistic theory developed by \citet{1980A&A....86..121S}. In this model, a fraction of the photons from an input seed spectrum is upscattered into a power-law component. We therefore modified the single-component Model~3 by including {\sc simpl}, which takes the accretion disc {\sc diskpbb} model as an input seed spectrum ({\sc simpl}$\otimes${\sc diskpbb} in XSPEC notation, where $\otimes$ denotes convolution). This modified model, later referred to as Model 3a, provides an acceptable fit quality ($\chi^{2}$/dof = 901/901), as also demonstrated by the residuals (Fig.~\ref{fig:model3ab}). The best-fitting parameters are listed in Table~\ref{tab:model3}.

As emphasized by \cite{2009PASP..121.1279S}, {\sc simpl} for simplicity does not include  any high-energy cutoff. However, any physical thermal Comptonization model should have a cutoff at photon energies higher than $kT_{\rm e}$. In order to constrain a possible high-energy attenuation, we added a multiplicative exponential cutoff model {\sc spexpcut} to the convolution model {\sc simpl$\otimes$diskpbb} (Model 3b in Table~\ref{tab:model3}). The model provides acceptable fit statistics ($\chi^{2}$/dof = 891/901), yielding a much harder index $\Gamma=1.8\pm0.1$ and lower scattered photon fraction $f_{\rm scat}$, and suggests the presence of a high-energy cutoff at $E_{\rm cut}=7.5_{-0.5}^{+0.8}$~keV. The improvement in the fit quality is, however, modest, $\Delta \chi^2\sim10$ for one fewer degree of freedom, implying that a high-energy cutoff is not strongly demanded by the data (the F-test probability of $2\times10^{-3}$ for null-hypothesis Model 3a). Additionally, we checked that replacing the {\sc diskpbb} model by {\sc diskbb} gives similar results.

Next, we replaced {\sc simpl} with the more detailed Comptonization model {\sc nthcomp}, where the electron temperature is a free parameter \citep{1987ApJ...319..643L,1996MNRAS.283..193Z,1999MNRAS.309..561Z}. The {\sc nthcomp} model approximates the Comptonization spectrum with the solution of the Kompaneets equation applying a relativistic correction to the energy transfer between photons and electrons. The input seed photons can be either blackbody or originating from a multi-colour accretion disc. The list of model parameters includes the asymptotic power-law photon index $\Gamma_{\rm nthcomp}$, the coronal electron temperature $kT_{\rm e,\ nthcomp}$, the seed photon temperature (linked to the inner accretion disc temperature in our case), and the normalization $N_{\rm nthcomp}$. Fitting by {\sc nthcomp} in combination with a $p$-free ``slim'' disc model {\sc diskpbb} provides a good fit with $\chi^{2}$/dof = 886/900. The model is not sensitive to the high-energy rollover, implying $kT_{\rm e,\ nthcomp}>100$~keV. An alternative combination of the {\sc nthcomp} component with a  standard accretion disc {\sc diskbb} component constrains the electron  temperature to $kT_{\rm e, nthcomp}=17.7_{-2.4}^{+1.3}$~keV, at comparable fit statistics ($\chi^{2}$/dof = 887/901) for one fewer free parameter. The best-fitting model parameters are listed in Table~\ref{tab:nthcomp}.

\begin{table}
\noindent
\centering
\caption{Same as Table~\ref{tab:simple}, but for Models 6a and 6b. }\label{tab:nthcomp}
\centering
\vspace{1mm}
 \begin{tabular}{|c|c|r|r|r|c|c|}
\hline\hline
Parameter & Units &  Value \\
\hline
\multicolumn{3}{c|}{Model 6a: {\sc const$\times$(diskpbb+nthcomp)}} \\
\hline
Constant& & $1.01\pm0.05$ \\
$kT$ & keV &   $0.93_{-0.03}^{+0.06}$ \\
$p$ &  & $0.80_{-0.07}^{+0.03}$ \\
$N_{\rm diskpbb}$ &  & $0.44_{-0.06}^{+0.03}$ \\
$R\ {\rm cos}^{1/2}i$  & km & $54_{-4}^{+2}$ \\
$\Gamma_{\rm nthcomp}$& & $3.39_{-0.13}^{+0.07}$  \\
$kT_{\rm e,\  nthcomp}$ & keV &   $365_{-220}^{+100}$ \\
$N_{\rm nthcomp}$& $10^{-3}$ & $1.9\pm0.3$  \\
$\chi^{2}$/dof& &  886/900 \\
\hline
\multicolumn{3}{c|}{Model 6b: {\sc  const$\times$(diskbb+nthcomp)}}\\
\hline
Constant& & $1.01\pm0.05$ \\
$kT$ & keV &   $0.97\pm0.04$ \\
$N_{\rm diskbb}$ &  & $0.40\pm0.04$ \\
$R\ {\rm cos}^{1/2}i$ & km & $52\pm3$ \\
$\Gamma_{\rm nthcomp}$& & $3.2\pm0.2$  \\
$kT_{\rm e,\  nthcomp}$ & keV &   $17.7_{-2.4}^{+1.3}$ \\
$N_{\rm nthcomp}$& $10^{-3}$ & $1.5\pm0.3$  \\
$\chi^{2}$/dof&  & 887/901  \\
\hline
\end{tabular}
\end{table}

\begin{figure}
\includegraphics[width=\columnwidth]{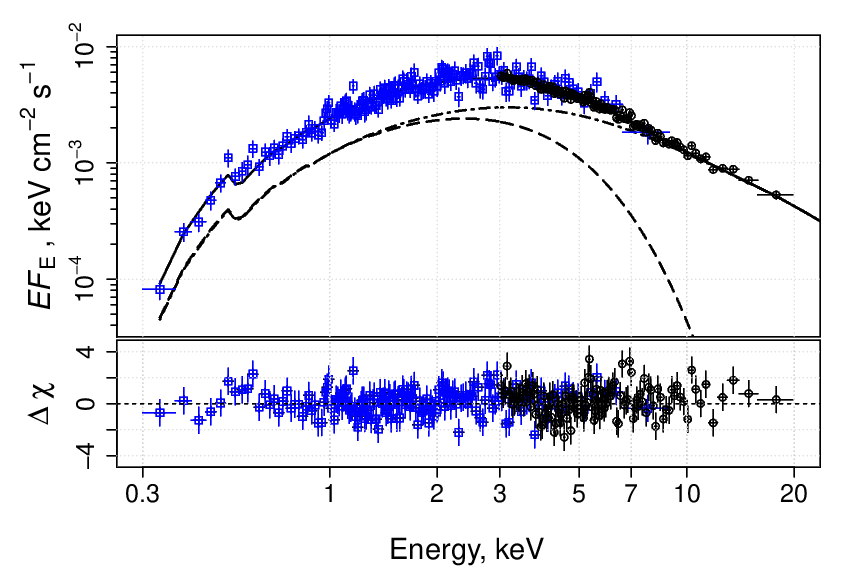}
\caption{Broad-band {\it Swift}/XRT and \nustar\ spectrum of \x8, shown by blue squares and black circles, respectively. The best-fitting spectral Model~6a ({\sc diskpbb + nthcomp}, Table~\ref{tab:nthcomp}) is shown by the solid black curve. The single accretion disc component {\sc diskbb} and the Comptonization {\sc nthcomp} component are shown by the dashed and dash-dotted curves, respectively. The bottom panel shows the fit residuals.}
\label{fig:model6b}
\end{figure}

\section{Discussion and conclusions}
\label{sec:discussion}

We have analyzed the hard X-ray (3--20~keV) spectrum of \x8\ based on the data obtained in 2017 with \nustar\ in the framework of the Legacy Program. In order to extend the spectrum into the softer X-ray band, we made use of the high-quality spectrum acquired with \textit{Swift}/XRT\ during the long observation of \x8\ in 2012.

In 2017, the X-ray flux of \x8\ in the 0.3--10~keV energy band (as found by extrapolating the \nustar\ spectrum, using Model~3a, to lower energies) was $1.8\times10^{-11}$\flux, which falls into the range of fluxes, $(1.1-2.0)\times10^{-11}$~\flux, observed from this source over a period of 16 years (1998--2014) with  {\it Swift}/XRT and other missions \citep{2015A&A...580A..71L}. This confirms that \x8\ is a relatively stable source compared to some other ULXs. Applying a line-of-sight absorption correction of $\sim$15 per cent \citep[inferred from the Galactic value $N_{\rm H}=1.1\times10^{21}$~cm$^{-2}$][]{2005A&A...440..775K}, to the observed flux, we find that the 0.3--10~keV isotropic luminosity of \x8\ is $1.6\times10^{39}$~\lum\ (for a distance of 817~kpc). The corresponding 3--20~keV luminosity $5.8\times10^{38}$~\lum\ is almost unaffected by absorption and has been directly measured for \x8\ for the first time. The broad-band, 0.3--20~keV, luminosity is $2.2\times10^{39}$~\lum\ and can be regarded as bolometric in X-rays, because the spectrum falls rapidly above 10~keV. We thus conclude that about 30 per cent of the total luminosity is radiated at energies above 10~keV. 

The broad-band (0.3--20~keV) spectrum of \x8\ can be well represented by a sum of two distinct components. The first one is a blackbody-like emission with a temperature $\sim 1$~keV, which presumably originates in an accretion disc. The second component, detected reliably up to 20~keV, dominates the emission above 10~keV. This hard X-ray tail is well fitted by a power law with a photon index $\Gamma\approx 3$. 

In the slim-disc interpretation often used to describe ULX spectra, we find the temperature gradient parameter $p\simeq0.75$ to be perfectly consistent with the value expected for a standard geometrically thin disc (Models 3a/b and 6a, see Tables~\ref{tab:model3} and \ref{tab:nthcomp}). Note that a single slim-disc model does not fit the broad-band spectrum, putting the temperature gradient at its hard limit $p=0.5$ (Model 3, Table~\ref{tab:simple}). This resembles the results obtained by \citet{2015A&A...580A..71L} using the narrow-band {\it Swift}/XRT data, who found $p=0.60\pm0.02$.  We conclude that the broad-band (0.3--20~keV) spectrum of \x8\ does not require a slim-disc interpretation. The normalization of the disc component varies in the wide range 0.4--1.4 between the models, implying the inner disc radius, $R\cos^{1/2}i$, to range between 50 and 100~km, broadly consistent with the compact object being a stellar-mass black hole. 

The hard tail can be well approximated by the empirical Comptonization model {\sc simpl}, in which a fraction of seed photons from the accretion disc are scattered  into a power-law component. Alternatively, the hard tail can be approximated by more accurate Comptonization models such as {\sc nthcomp}. In both cases, introducing a high-energy cutoff leads to a moderate improvement in the goodness of fit, although the statistical significance of this result is not high. 

Overall, in its broad-band spectral properties \x8\ resembles black-hole X-ray binaries in their  very high (or steep power-law) state. Specifically, the characteristic temperature of the accretion disc is $\sim 1$~keV, the hard tail is steep ($\Gamma\sim 3$) and contributes significantly ($\gtrsim 30$\%) to the total luminosity. Furthermore, a number of X-ray binaries have been caught at luminosities $\sim 10^{39}$\lum\ in their very high state \citep[e.g.][]{2006csxs.book..157M,2007A&ARv..15....1D}, similar to the observed luminosity of \x8. All this suggests that \x8\ may be a black-hole system accreting at a high but sub-critical rate, in contrast to higher luminosity ULXs, where super-Eddington accretion presumably takes place.

As mentioned in Section~\ref{sec:intro}, the recent discovery of X-ray pulsations from a number of ULXs has demonstrated that in such objects super-Eddington accretion proceeds onto a magnetized neutron star, rather than a black hole. Can \x8\ too be an object of this type?

As demonstrated by \cite{2018ApJ...856..128W}, the spectra of ULX-pulsars are similar to those of the majority of well-studied ULXs. Specifically, they peak (when plotted in $EF_E$ units) at 5--10~keV and can be described by a combination of stable thermal emission from an accretion disc and a pulsating non-thermal component that  is well fit by a cutoff power law with $\Gamma\sim 0$--1 and $E_{\rm cut}\sim 5$--20~keV. The latter presumably originates in accretion columns at the surface of the neutron star.  The spectrum of \x8, studied here, can be described by a similar combination of thermal and non-thermal components, with the latter having best-fitting parameters ($\Gamma\sim 1$ and $E_{\rm cut}\sim 5$~keV) similar to those of ULX-pulsars. However, the contribution of this hard component to the total spectrum is much weaker than in ULX-pulsars. Furthermore, the thermal component, presumably associated with the accretion disc, is softer. As a result, the spectrum of \x8\ peaks at a much lower energy, $\sim 2.5$~keV, and looks quite different from the spectra of ULX-pulsars and from those of the majority of ULXs. This suggests that \x8\ is {\sl not} a neutron-star accretor.

Nevertheless, if we interpret the soft component as emission from the accretion disc truncated by a neutron star's magnetic field, the inferred size of the emission region implies that the magnetosphere extends to $\gtrsim 5$ stellar radii. In this case, we would expect an additional harder component coming from the accretion columns at the neutron star surface with luminosity several times higher than the observed soft disc luminosity. Furthermore, the accretion rate will then be a few times higher than that implied by the observed luminosity, i.e. highly supercritical. The actually observed luminosity of the hard component in the spectrum of \x8\ is, however, less than half of the total one, which poses a serious problem for the neutron star scenario. It is possible though that the hard X-ray component is weak because the accretion columns are partially obscured by the inner thick disc. In that case, there would be no X-ray pulsations either; and indeed, our and previous \citep{2011MNRAS.417..464M} searches for pulsations in \x8\ have been unsuccessful.

In our view, the interpretation of the data with the model invoking a black hole is preferred.
Based on the similarity with the very high state of black-hole X-ray binaries, we may expect the hard spectral tail of \x8\ to extend well above 20~keV as, for example, is seen in GRS~1915+105 \citep{2001ApJ...554L..45Z}. This could be verified  with specially dedicated longer \nustar\ observations.

\section*{Acknowledgments}

This work has made use of data from the \nustar\ mission, a project led by the California Institute of Technology, managed by the Jet Propulsion Laboratory and funded by NASA, and observations obtained with \textit{XMM-Newton}, an ESA science mission with instruments and contributions directly funded by ESA Member States and NASA. The research has made use of the \nustar\ Data Analysis Software jointly developed by the ASI Science Data Center (ASDC, Italy) and the California Institute of Technology. The data reduction and spectral analysis were performed by RK and SS with the support of grant 14-12-01315 from the Russian Science Foundation. SST and JP took part in the interpretation of the results with the support of  grant 14.W03.31.0021 from the Ministry of Education and Science of the Russian Federation.



\bsp	
\label{lastpage}
\end{document}